\documentclass[a4paper,11pt]{article}
\pdfoutput=1 

\usepackage{jcappub} 
\usepackage{subcaption}
\usepackage[T1]{fontenc} 
\usepackage{sidecap}
\usepackage{subcaption}
\usepackage{booktabs}
\usepackage{array}
\usepackage{cleveref}
\usepackage{xcolor}
\usepackage{soul}
\usepackage{amsmath, graphicx}
\usepackage{diagbox}

\title{Defying eternal inflation in warm inflation with a negative running}

\author[a,1]{Gabriele Montefalcone,\note{Corresponding author.}}
\author[b]{Rudnei O. Ramos,}
\author[c]{Gustavo S. Vicente}
\author[a,d,e]{and Katherine Freese}

\affiliation[a]{Texas Center for Cosmology and Astroparticle Physics, Weinberg Institute for Theoretical Physics, Department of Physics, University of Texas, Austin, Texas 78751, USA}
\affiliation[b]{Departamento de F\'{\i}sica Te\'orica, Universidade do Estado do Rio de Janeiro, 20550-013 Rio de Janeiro, RJ, Brazil}
\affiliation[c]{Faculdade de Tecnologia, Universidade do Estado do Rio de Janeiro, 27537-000 Resende, RJ, Brazil}
\affiliation[d]{The Oskar Klein Centre, Department of Physics, Stockholm University, AlbaNova, SE-10691 Stockholm, Sweden}
\affiliation[e]{Nordic Institute for Theoretical Physics (NORDITA), 106 91 Stockholm, Sweden}

\emailAdd{montefalcone@utexas.edu}
\emailAdd{rudnei@uerj.br}
\emailAdd{gustavo@fat.uerj.br}
\emailAdd{ktfreese@utexas.edu}

\abstract{ It was pointed out previously~\cite{Kinney:2014jya} that a
sufficiently negative running  of the spectral index of curvature
perturbations from (ordinary i.e. cold) inflation is able to prevent eternal inflation from
ever occurring. Here, we reevaluate those original results, but in the
context of warm inflation, in which a substantial  radiation component (produced by the inflaton) exists throughout the inflationary period.
 We demonstrate that the same general
requirements 
found in the context
of ordinary (cold) inflation also hold true in warm inflation; indeed an even tinier amount of negative running
is sufficient to prevent eternal inflation. This is particularly pertinent, as models featuring negative running are more generic in warm inflation scenarios.  
Finally, the condition for the existence of eternal inflation in cold inflation --- that the curvature perturbation amplitude exceed unity on superhorizon scales --- becomes more restrictive in the case of warm inflation. The curvature perturbations must be even larger, i.e. even farther out on the potential, away from the part of the potential where observables, e.g. in the Cosmic Microwave Background, are produced.}

\preprint{UTWI-40-2023, NORDITA-2023-067}

\begin{document}
\maketitle

\section{Introduction}
Inflation~\cite{Guth:1980zm, Linde:1981mu, Albrecht:1982wi, Kazanas:1980tx, Starobinsky:1980te, Sato:1980yn, Mukhanov:1981xt, Linde:1983gd, Mukhanov:1990me}, a brief period of accelerated expansion in the early universe, is currently the most-compelling framework to produce the large scale homogeneity, isotropy, and oldness of the Universe, as well as provide a mechanism for generating the density fluctuations that ultimately give rise to the observed large-scale
structures.  

A large class of inflaton models involves a single scalar field slowly rolling down a nearly
flat potential, inducing a quasi-de Sitter phase. Once the field nears the minimum of its potential, the accelerated expansion ends and the inflaton energy converts to
radiation, thereby reheating the Universe.
Hereafter, we refer to this standard scenario as cold inflation (CI) to emphasize the absence of substantial interactions between the inflaton field and other degrees of freedom during the inflationary period.

A well-established alternative framework to this standard picture is warm inflation (WI)~\cite{Berera:1995ie} (see
also Refs.~\cite{Berera:2008ar,Bastero-Gil:2009sdq,Kamali:2023lzq} for
reviews), where instead the inflaton field  is thermally coupled to a bath of radiation, continuously sourcing its production throughout the accelerated expansion. 
Via these
dissipative effects, WI is able to smoothly route the Universe to
a radiation dominated era, alleviating the need for a separate
reheating phase. 
The presence of dissipation and a radiation bath not only alter the inflaton dynamics but also its perturbations, which now become predominantly thermal in origin, with quantum fluctuations being subdominant in the limit of a
large dissipation rate between the two sectors. For a given inflaton
potential, this leads to significantly different predictions, with respect to those obtained in CI, for many
primordial observables, such as the scalar spectral index $n_s$ and
the tensor-to-scalar ratio $r$~\cite{Bartrum:2013fia,Bastero-Gil:2014raa,Bastero-Gil:2016qru,Bastero-Gil:2019gao}. These divergent predictions provide a means to distinguish between a CI and WI scenario and also allow the WI framework to reconcile some of the simplest inflaton potentials, otherwise excluded by data in the CI context~\cite{Benetti:2016jhf}.

In addition WI constructions can naturally overcome some of the issues in the standard CI picture. Specifically, the decay of the inflaton into radiation during inflation acts as a source of (thermal) friction, which allows 
WI models to occur for inflaton masses above the Hubble scale and/or with a subplanckian field excursion. This unique ingredient of the WI framework can provide a solution to the
$\eta$-problem~\cite{Bastero-Gil:2019gao}, as well as make WI
consistent with the recently proposed Swampland Conjectures in String
Theory~\cite{Das:2018hqy,Motaharfar:2018zyb,Das:2018rpg,Das:2019hto,Kamali:2019xnt,Brandenberger:2020oav,Das:2020xmh}, generally in strong
tension with the slow-roll CI scenario.\footnote{For a a detailed discussion on how WI is able to satisfy these different conditions, see e.g. Ref.\cite{Berera:2023liv}. }

Another important and generic feature of inflationary models is \textit{eternal inflation}
(EI)~\cite{Vilenkin:1983xq,Guth:1985ya,Linde:1986fc,Linde:1986fd,Starobinsky:1986fx},
which happens when the stochastic fluctuations of the inflaton field
dominate over its deterministic evolution. {}For the regions where the
field is systematically deviated {\it upwards},   a self-reproduction
regime (SRR) of causally disconnected Hubble-sized regions will take
place and inflation never ends. In CI, the stochastic fluctuations are
only quantum in origin, while in WI both quantum and thermal
fluctuations may be responsible for the onset of EI. However, in WI
one also needs to account for the effect of the dissipation on the
deterministic field excursion and the scale of the inflaton
potential. In Ref.~\cite{Vicente:2015hga}, by analyzing some physically motivated WI constructions, it was demonstrated  that the dissipative and thermal characteristics of WI offer a mechanism for mitigating the onset of EI,
thereby suggesting that EI may be prevented altogether in a strongly dissipative regime.\footnote{A detailed analytical and numerical study of the onset of EI in the strongly dissipative regime in fact confirms this intuition, i.e., WI in a strongly dissipative regime can suppress the emergence of EI altogether~\cite{MRVFeistudy}.}.

In this work, we aim to investigate this statement more generally and determine the requirements for the onset of EI in a WI setting, in terms of a bound on the amplitude of the curvature power spectrum and its functional form. In fact, in the previous work of Ref.~\cite{Kinney:2014jya}, it was shown, in
the context of CI, that there is a connection between the value of the
running of the scalar spectral index $\alpha_s$ of the curvature
perturbations, and the onset of an eternally inflating regime.  Specifically, EI is prevented if
$\alpha_s$ is sufficiently negative. 

In the following, our main objective is to extend the bound obtained in
Ref.~\cite{Kinney:2014jya} to the WI paradigm and investigate
whether such new bound on $\alpha_s$, together with the obtained
values of $\alpha_s$ in physically motivated WI constructions~\cite{Benetti:2016jhf,Das:2022ubr},
prevents EI  and is consistent with the latest observational constraints from Planck~\cite{Planck:2018jri}. We will show that in the WI framework an even tinier amount of negative running is sufficient to prevent EI, and provide examples of several physically motivated WI constructions that produce a running of the spectral index that is sufficiently negative to prevent EI from ever occurring, while being consistent with CMB observations. 
Note that this is a nontrivial issue in WI, as there are four competing effects:   (1) in WI there is a reduced deterministic
field excursion due to the friction caused by the dissipation of the
inflaton field into the radiation. At the same time,  (2) there is an
enhancement of the stochastic fluctuations of the inflaton field due
to the addition of classical thermal fluctuations.  Both of these
effects act in a way to facilitate the onset of EI.  On the other hand, (3) the inflaton fluctuations are also dumped into the radiation bath at
a rate proportional to the strength of the dissipation.   In addition, (4) the
height of the inflaton potential, which sets the scale of the
stochastic fluctuations, is suppressed relative to the CI case, such
as to counteract the large thermal/dissipation enhancement factor in
the scalar power spectrum in WI and reproduce the observed density
perturbations. As we are going to explain and show, these latter effects surpass
the former ones, which together with a smaller negative prediction for
the running, acts to prevent EI altogether in WI.
 
The paper is organized as follows.  In Sec.~\ref{EICI}, we briefly
review the relationship between EI and the
amplitude of the curvature perturbation spectrum in the CI scenario, and introduce the Starobinsky stochastic formalism that more easily generalizes to the WI case.  The stochastic (diffusion) and deterministic
(drift) terms are discussed along with the definition of the
fluctuation dominated regime (FDR).  The FDR condition is then derived
and expressed in terms of the scalar power spectrum.  In
Sec.~\ref{EIWI}, we extend the discussion to the case of WI, where the
diffusion and drift terms now incorporate both quantum and thermal
effects.  A generalized FDR condition is derived in terms of the WI
scalar power spectrum.  Our main results start from Sec.~\ref{upper},
where we present the power spectrum in terms of the spectral index and
its running, and derive the upper bound on $\alpha_s$ necessary in
order to prevent EI.  We show some examples in order to compare CI and
WI results regarding this upper bound.  We then also discuss the
higher order correction terms, i.e., the running of the running of the
spectral index, $\beta_s$, and the successive higher order terms
$\gamma_s$, $\delta_s$, etc, and their impact in the $\alpha_s$ upper
limit.  Our final considerations and conclusions are presented in
Section~\ref{conclusions}.

\section{Eternal Inflation Regime in Standard Cold Inflation }
\label{EICI}

In this section we review the conditions required for producing an eternal inflation (EI) regime in the case of standard CI. We begin with a simple comparison between quantum fluctuations of the rolling field with its classical evolution. Then we reproduce the same criterion using the Starobinsky stochastic formalism~\cite{Starobinsky:1986fx} that more easily generalizes when we turn to the WI scenario.

In CI, eternal inflation occurs when the quantum fluctuations in the inflaton field dominate over the classical field evolution.  The amplitude of quantum fluctuations in the inflaton field during inflation is
\begin{equation}
\delta\phi_Q \equiv \left\langle \delta \phi^2\right\rangle^{1/2} = \frac{H}{2 \pi}.
\end{equation}
For EI to take place, this quantum fluctuation amplitude must be larger than the classical field variation over approximately a Hubble time ($\Delta t\sim
H^{-1}$),
\begin{equation}
\delta\phi_C = \frac{\dot\phi}{H}.
\end{equation}
Therefore, the condition for EI can be written as:
\begin{equation}
\frac{\delta\phi_Q}{\delta\phi_C} = \frac{H^2}{2 \pi \dot\phi} > 1.\label{eq:eternal}
\end{equation}
We note that the fraction (\ref{eq:eternal}) is identical to the amplitude of the curvature perturbation for modes crossing the horizon during inflation,
\begin{equation}
P\left(k\right) =\left(\frac{H^2}{2\pi\dot{\phi}}\right)^2. \label{eq:DeltaR2_CI}
\end{equation}
In other words, the condition for EI in CI is that the curvature perturbations amplitude must
exceed unity~\cite{Goncharov:1987ir,Guth:2007ng,Kinney:2014jya},
\begin{equation}
\label{eq:FDR_CI}
P\left(k\right) > 1 \,\,\,\,\,\,\,   ({\rm 
 CI}) \, .
\end{equation}
This makes sense since in the standard CI scenario the
curvature perturbations are equal to the amplitude of the quantum fluctuations in the inflaton,
in units of the field variation in a Hubble time.

{\it Starobinsky formulation:} In order to be able to generalize these results  beyond the case of the standard CI models, 
we show a derivation of the results in the CI case
by expressing the dynamics of the inflaton field
in terms of the Starobinsky stochastic inflation
program.  This approach provides a simple way for
describing the backreaction of the  short-wavelength inflaton modes on
the dynamics of the long-wavelength ones. The equation for the
homogeneous inflaton field $\phi$ is written as a Langevin-like
equation of the form:
\begin{eqnarray}
  \dot{\phi}=-f(\phi)+\sqrt{2D^{(2)}(\phi)}\zeta_q,
  \label{eq:EOM_cold}
\end{eqnarray}
where
\begin{eqnarray}
    f(\phi)\equiv \frac{V_{,\phi}(\phi)}{3H(\phi)} \quad , \quad
    D^{(2)}(\phi)=\frac{H^3(\phi)}{8\pi^2},
\end{eqnarray}
are the drift and diffusion coefficients, respectively, while
$\zeta_q(t)$ is a Gaussian noise term that accounts for the stochastic
(quantum) fluctuations of the inflaton field, with correlation
function
$\langle\zeta_q(t)\zeta_q(t^\prime)\rangle=\delta(t-t^\prime)$.

The presence of an eternally inflating patch requires that the inflationary dynamics goes through a fluctuations dominated regime (FDR). This is equivalent to say that EI will occur when the stochastic fluctuations of the inflaton
field, $\delta\phi_S$, dominate over its deterministic evolution,
$\delta\phi_D$, over approximately a Hubble time,
\begin{eqnarray}
\delta\phi_S>\delta\phi_D, 
\label{eq:FDR1}
\end{eqnarray}
and where the deterministic and stochastic field excursions over
approximately a Hubble time are given, respectively, by
\begin{align}    &\delta\phi_D=\frac{|f(\phi)|}{H(\phi)},
  \label{eq:deltaphiD_C}\\
  &\delta\phi_S=\sqrt{\frac{2D^{(2)}(\phi)}{H(\phi)}}.
  \label{eq:deltaphiS_C}
\end{align}
This then leads to the condition for the presence of a FDR,
\begin{eqnarray}
 \sqrt{\frac{2D^{(2)}(\phi)}{H(\phi)}}>
 \frac{|f(\phi)|}{H(\phi)}. \label{eq:FDR}
\end{eqnarray}
The above condition emphasizes that when the diffusion term dominates
over the drift one, the time evolution of the inflaton field is
strongly nondeterministic.  In CI, the FDR condition
reads\footnote{Note, in the following we took $|f(\phi)|=\dot{\phi}$,
which is true in the deterministic (classical) regime, where we ignore
the stochastic fluctuations.}
\begin{eqnarray}
\frac{H^2}{2\pi\dot{\phi}}>1.
\end{eqnarray}
which matches the amplitude of the scalar curvature perturbations, Eq.\eqref{eq:DeltaR2_CI}. Hence, again we find the result of Eq.~\eqref{eq:FDR_CI} above, that the condition for EI in CI is that the curvature
perturbations amplitude must exceed
unity~\cite{Goncharov:1987ir,Guth:2007ng,Kinney:2014jya},
$P (k) > 1$. 

In the standard CI scenario presented above, the stochastic fluctuations and hence the curvature perturbations were determined solely by the quantum fluctuations of the inflaton field. 
This is, however, not the case in WI, where classical thermal fluctuations in the inflaton field act as an additional source of
curvature perturbations, generally dominating over the quantum fluctuations. In this sense, we also expect that the FDR condition in
the WI context  will depart from its simple form as given by
Eq.~\eqref{eq:FDR_CI}.  We explore this statement explicitly in the next section.

\section{Eternal Inflation Regime in Warm Inflation}
\label{EIWI}

We now turn to the condition for eternal inflation in the case of WI. We introduce the inflaton field equation of motion in WI expressed in terms of the Starobinsky stochastic program, placing particular emphasis to the effects of thermal dissipation on the inflaton dynamics. We then use this formalism to derive the generalized condition for EI that is valid when temperature and dissipation effects become relevant. 

The equivalent Langevin-like equation of motion of the inflaton field
$\phi$ in WI reads~\cite{Vicente:2015hga}:
\begin{eqnarray}
  \dot{\phi}=\frac{1}{3H+\Upsilon}\left(-V_{,\phi}(\phi)+\xi_q+\xi_T\right), \label{eq:EOM_warm}
\end{eqnarray}
where, in contrast to CI, we included a non-negligible dissipation rate $\Upsilon\equiv \Upsilon(\phi,T)$, which accounts for the energy transfer between the inflaton field and the radiation bath (at temperature $T$) present during inflation.\footnote{The detailed form of the dissipation rate $\Upsilon$ is model-dependent and is generically a function of both the inflaton field $\phi$ and temperature $T$ of the radiation bath. For explicit constructions of dissipation rates, we refer the reader to~\cite{Berera:2002sp, Moss:2006gt, Berera:2008ar,Bastero-Gil:2010dgy,Bastero-Gil:2012akf,Bastero-Gil:2016qru, Bastero-Gil:2019gao, Berghaus:2019whh}. } In addition, we also introduced two stochastic Gaussian noise
terms, $\xi_q$ and $\xi_T$, which account for  the quantum and thermal fluctuations, respectively,
with two-point correlation functions given
by~\cite{Ramos:2013nsa,Kamali:2023lzq,Montefalcone:2023pvh}:
\begin{align}
    \label{xiT}
    &\langle
    \xi_{T}(\mathbf{x},t)\xi_{T}(\mathbf{x^\prime},t^\prime)\rangle
    =\frac{6HQT}{a(t)^3}\delta^3(\mathbf{x}-\mathbf{x}^\prime)
    \delta(t-t^\prime), \\
    \label{xiq}
    &\langle
    \xi_{q}(\mathbf{x},t)\xi_{q}(\mathbf{x^\prime},t^\prime)\rangle
    =\frac{H^2
      \sqrt{9+12\pi Q} (1+2n_{*})}{\pi a(t)^3}
    \delta^3(\mathbf{x}-\mathbf{x}^\prime) \delta(t-t^\prime).
\end{align}
Here, $n_*$ denotes the possible inflaton statistical distribution due
to the presence of the radiation bath, generally assumed to be the
equilibrium Bose-Einstein distribution, i.e., $n_* = [\exp(H/T) -
  1]^{-1}$; and $Q$ is the dimensionless ratio measuring the effectiveness at which the inflaton converts into radiation, defined as:
  \begin{equation}
      Q\equiv \frac{\Upsilon}{3H}.
  \end{equation}
 For $Q\gg 1$, a strongly
dissipative WI (SDWI) regime is achieved, while $Q<1$ represents the
weak dissipative WI (WDWI) regime. In all cases, WI requires $T>H$, which is roughly the criterion for which thermal fluctuations dominate over quantum fluctuations~\cite{Berera:1995ie}.
We note that for $Q\ll1$ and
$T<H$, the CI limit is  recovered.

For the purpose of this work, we are interested in the evolution of the nearly homogeneous
inflaton field inside a Hubble volume and over a Hubble time. Hence, we must integrate out the
spatial Dirac-delta function from the correlation functions. To do so,
one simply notes that $\delta^{3}(\mathbf{x}-\mathbf{x^\prime})$
corresponds to an inverse volume factor. The natural volume to be
taken is the de Sitter volume of the horizon, $V_H\equiv
\frac{4\pi}{3}\left(1/H\right)^3$, such that we can simply take
$\delta^{3}(\mathbf{x}-\mathbf{x^\prime})\rightarrow
1/V_H$. Additionally, to recover the standard CI result,
in the denominator we normalize the scale factor $a$, such that $a(t=1/H)=1$.  With these substitutions, Eq.~\eqref{eq:EOM_warm} becomes:
\begin{align}
    \dot{\phi}&=-f_w(\phi)+\sqrt{2D^{(2)}_{\rm{vac}}}\,\zeta_q
    +\sqrt{2D^{(2)}_{\rm{diss}}}\,\zeta_T,
\label{WIlangevin}
\end{align}
where $\zeta_q$ and $\zeta_T$ are now the quantum and dissipation
noises with the correlation functions  given by $\langle
\zeta_i(t)\zeta_i(t^\prime)\rangle= \delta(t-t^\prime)$, for $i=\{ q,
T\}$.  In Eq.~(\ref{WIlangevin}), we also have that $f_w(\phi)$ is the
drift coefficient, while $D^{(2)}_{\rm{vac}}$ and
$D^{(2)}_{\rm{diss}}$ are the quantum and thermal contributions to the
diffusion coefficient, respectively.  Their expressions are given by
\begin{align}
  &f_w(\phi)\equiv\frac{V_{,\phi}(\phi)}{3H(1+Q)}, \label{eq:driftC_warm}
  \\
  &D^{(2)}_{\rm{vac}}\equiv\frac{H^{3}}{8\pi^2}\frac{1+2n_*}{(1+Q)^2}
  \sqrt{1+\frac{4\pi
      Q}{3}}, 
\label{eq:diffusionC_warmq}
\\ & D^{(2)}_{\rm{diss}}\equiv \frac{H^{3}}{8\pi^2} \frac{2\pi
  Q}{(1+Q)^2} \,\frac{T}{H}. \label{eq:diffusionC_warmT}
\end{align}
The deterministic evolution and the stochastic fluctuations of the
inflaton field now become\footnote{To obtain
Eq.~\eqref{eq:deltaphiS_W}, we assumed statistical independence
between the thermal and quantum noises, i.e.,
$\langle\zeta_T\zeta_q\rangle=0$, such that the effective diffusion
coefficient
$D^{(2)}_w=D^{(2)}_{\rm{vac}}+D^{(2)}_{\rm{diss}}$~\cite{Ramos:2013nsa}.}
\begin{align}
  &\delta\phi^{w}_D=\frac{|f_w(\phi)|}{H},\label{eq:deltaphiD_W}
  \\
  &\delta\phi^{w}_S=\sqrt{\frac{2\left(D^{(2)}_{\rm{vac}}+D^{(2)}_{\rm{diss}}
      \right)}{H}}. \label{eq:deltaphiS_W}
\end{align}

Combining Eqs.~\eqref{eq:driftC_warm}-\eqref{eq:deltaphiS_W}, the FDR
condition, Eq.~\eqref{eq:FDR1}, in the WI context can now be expressed
as
\begin{eqnarray}
    \frac{H}{2\pi
      \dot{\phi}}>\frac{1+Q}{\sqrt{(1+2n_*)\sqrt{1+\frac{4\pi
            Q}{3}}+2\pi Q \,\frac{T}{H}\, }}. \label{eq:FDR_warmv0}
\end{eqnarray}
As previously mentioned, in WI the addition of
thermal effects has also a significant impact on the form of the primordial power spectrum which in WI is given by~\cite{Kamali:2023lzq}:
\begin{eqnarray}
    P(k)=\left(\frac{H^{2}}{2 \pi \dot{\phi}}\right)^2\left(1 + 2
    n_{*}+\frac{2 \sqrt{3} \pi Q}{\sqrt{3+4 \pi Q}} \frac{T}{H}\right)
    G(Q).\label{eq:DeltaR_warm}
\end{eqnarray}
where the multiplicative factor $G(Q)$ accounts for the 
direct coupling of the inflaton and radiation fluctuations due to a temperature-dependent dissipation rate, and can only be determined numerically by solving the full set of perturbation equations~\cite{Graham:2009bf, Bastero-Gil:2011rva,Montefalcone:2023pvh}. For a dissipation rate $\Upsilon\propto T^c$, if $c=0$ we have $G(Q)=1$, while $G(Q)$ is larger (smaller) than one for $c >0$ ($c<0$).  

 We can now use Eq.\eqref{eq:DeltaR_warm} and rewrite the FDR condition in WI, Eq.~\eqref{eq:FDR_warmv0}, in terms of the amplitude of the curvature perturbations, similarly to what we previously did for the CI case. In formulas, the condition for EI in WI is
\begin{eqnarray}
    P(k)> P_{\rm{FDR}} \,\,\,\,\,\,\,   ({\rm 
 WI}) \, ,\label{eq:EIC_warm}
\end{eqnarray}
where
\begin{eqnarray}
    P_{\rm{FDR}}\equiv (1+Q)^2\frac{\left(1 + 2 n_{*}+\frac{2 \sqrt{3}
        \pi Q}{\sqrt{3+4 \pi Q}}
      \frac{T}{H}\right)G(Q)}{(1+2n_*)\sqrt{1+\frac{4\pi Q}{3}}+2\pi Q\,\frac{T}{H}\, }. \label{eq:P_FDR}
\end{eqnarray}
$P_{\rm{FDR}}$ defines the maximum amplitude of the curvature perturbations in order to avoid the onset of a fluctuations dominated regime. In other words, the condition for EI in WI is that the curvature perturbations amplitude must
exceed $P_{\rm{FDR}}$. 
Note that in the limit of $Q=0$, i.e. no dissipation,
Eq.~\eqref{eq:EIC_warm} reduces to the standard CI result: $P(k)>1$. For $Q>0$, we have that $P_{\rm{FDR}}>1$. More specifically, there
is a positive correlation between the dissipation strength $Q$ and
$P_{\rm{FDR}}$. This becomes evident in the SDWI regime, i.e. for
$Q\gg 1$. In this regime, we can approximate the scalar dissipation
function  $G(Q)$ with a power-law function of
$Q$~\cite{Montefalcone:2023pvh}, i.e.  $G(Q)\simeq a_G Q^{b_G}$, where both
the prefactor $a_G$ and the exponent $b_G$ are generically $\sim \mathcal{O}(1)$ numbers and additionally $b_G>0$ $(<0)$ for a positive (negative) temperature power
dependence of the dissipation rate.\footnote{The dissipation rate can be
generally parametrized as $\Upsilon\propto\,T^c$, where both $c>0$ (e.g.,
$c=1$~\cite{Bastero-Gil:2016qru} and $c=3$~\cite{Moss:2006gt}) and
$c<0$ (e.g., $c=-1$~\cite{Bastero-Gil:2019gao}) have been shown to
provide successful WI constructions.} Plugging this back into
Eq.~\eqref{eq:EIC_warm}, we obtain that
\begin{equation}
    P_{\rm{FDR}}(Q\gg 1)\simeq Q^2\frac{\sqrt{3\pi
        Q}\,\frac{T}{H}\,a_G Q^{b_G}}{2\pi Q
     \,\frac{T}{H}\,}  \simeq
    \sqrt{\left(\frac{3a_G^2}{4\pi}\right)}Q^{(2b_G+3)/2}.
    \label{eq:EIC_SDwarm}
\end{equation}
As long as $b_G>-1.5$ (which is true for all dissipation rates of
physical interest~\cite{Montefalcone:2023pvh}), there is a positive
power-law dependence on $Q$, such that $P_{\rm{FDR}}\gg1$ for $Q\gg
1$.

As a whole, the generalized condition for eternal inflation given by
Eq.~\eqref{eq:EIC_warm} clearly shows that in a WI setting, the onset of EI is suppressed relative to the standard CI case, and
this suppression is even more facilitated in the limit of strong dissipation. In other words, in WI the condition for the existence of EI is more restrictive than in the CI case: it is not sufficient for the amplitude of the curvature perturbations to exceed unity on superhorizon scales, it must be even larger ($>P_{\rm{FDR}}$), i.e. even farther
out on the inflaton potential, away from the part of the potential where observables, such as the Cosmic
Microwave Background, are produced.

This
conclusion may seem at first counter-intuitive since in the context of
WI we expect: (1) a reduced deterministic field excursion due to the
friction caused by the dissipation of the inflaton field into the
radiation and (2) an enhancement of the stochastic fluctuations of the
inflaton field due to the addition of classical thermal
fluctuations. Effects (1) and (2) both act to decrease
$\delta\phi_D$ and increase $\delta\phi_S$. Hence, both of these
effects contribute to enhance the ratio $\delta\phi_S/\delta\phi_D$,
thereby facilitating the onset of EI. This however ignores two
additional effects common to all WI constructions which suppress the
ratio $\delta\phi_S/\delta\phi_D$ and generally dominate over the
first two effects. Specifically, in WI: (3) the inflaton fluctuations
are dumped into the radiation bath\footnote{The dumping of the
inflaton fluctuations in the radiation bath is represented by the
factor of proportionality of $(1+Q)^2$ in the definition of
$P_{\rm{FDR}}$, Eq.~\eqref{eq:P_FDR}. } at a rate proportional to the
dissipation strength $Q$ and (4) the height of the inflaton potential,
which sets the scale of the stochastic fluctuations, is suppressed
relative to the CI case, such as to counteract the large thermal
enhancement factor in the power spectrum and to reproduce the observed
density perturbations~\cite{Montefalcone:2022owy}. Clearly, both
effects (3) and (4) decrease the resulting $\delta\phi_S$, thereby
suppressing the onset of EI as anticipated above\footnote{In the
forthcoming paper~\cite{MRVFeistudy}, an explicit derivation is
provided showing how each of the four aforementioned dissipative
effects impacts the stochastic and deterministic field excursion in
WI, as well as a direct comparison to the dynamics in the CI
scenario.}.

In the next section, we investigate this statement analytically. More
specifically, we extend the work of Kinney and Freese in
Ref.~\cite{Kinney:2014jya} and derive a generalized upper bound on the
running of the scalar spectral index to prevent EI to occur when in
the presence of thermal dissipation.

\section{An Upper Bound on the Running}
\label{upper}

The scalar power spectrum for curvature perturbations can be written
in terms of the scalar spectral index $n_s$ and of the running
$\alpha_s$, as
\begin{eqnarray}    P(k)=P_*\left(\frac{k}{k_*}\right)^{n_s-1+\frac{\alpha_s}{2}
    \ln(k/k_*)},
\end{eqnarray}
where $k_*$ is a pivot scale, which is typically the scale relevant
for CMB observations, i.e., $k_*=0.05h~\rm{Mpc}^{-1}$, for which the
amplitude of the scalar power spectrum is constrained to be $P_*\simeq
2.1 \times 10^{-9}$~\cite{Planck:2018vyg}. Additionally,
\textit{Planck} data also constrain the values of $n_s$ and $\alpha_s$
to be (at 68\% C.L.)~\cite{Planck:2018jri}\footnote{ Note, if the
running of the running $\beta_s$ is included in the analysis, the
result for $\alpha_s$ changes. Specifically, for the $\Lambda$CDM
model, the Planck 2018 TT(TT,TE,EE)+lowE+lensing  data has a slight
preference for a positive value of $\alpha_s$, but overall large
negative values for $\alpha_s$ are still allowed at the 95\%
C.L.~\cite{Planck:2018jri}.},
     \begin{align}
n_{\mathrm{s}} & =0.9641 \pm 0.0044, \\ \alpha_s & =-0.0045 \pm
0.0067. \label{eq:Planck_alphaS}
     \end{align}
While the value of the running $\alpha_s$ is still poorly constrained,
these CMB measurements clearly indicate that the spectral index is
red, i.e., $n_s-1<0$, with a 99.7\% confidence limit of approximately
$0.95<n_s<0.98$. {}For a constant red spectral index and no running,
i.e., $\alpha_s=0$, EI is inevitable since there is always a scale
$k_{\rm{FDR}}\ll k_*$ such that for all $k<k_{\rm{FDR}}$, the
condition for EI, Eq.~\eqref{eq:EIC_warm}, is satisfied.  Thus, as
emphasized in Ref.~\cite{Kinney:2014jya}, the simplest case of
interest is that of a constant negative running, $\alpha_s<0$.

Notice that a constant negative running means that the spectral index
gets redder on small scales $k \gg k_*$, and bluer on large scales, $k
\ll k_*$, such that for a sufficiently negative $\alpha_s$, the
spectral index for $k\ll k_*$ will eventually exceed unity, $n_s - 1 >
0$. This is sufficient to prevent EI as long as the scalar curvature
power spectrum $P (k)$ is bounded below $P_{\rm{FDR}}$ on large
scales,
\begin{eqnarray}
    \ln P(k)= \ln P_*
    +\left[n_s-1+\frac{\alpha_s}{2}\ln\left(\frac{k}{k_*}\right)\right]
    \ln\left(\frac{k}{k_*}\right)<\ln
    P_{\rm{FDR}},
\end{eqnarray}
for all $k<k_*$. In this scenario, the scalar power spectrum will have
an extremum at some wavenumber $k_{\rm{max}}<k_{*}$, given
by~\cite{Kinney:2014jya}
\begin{eqnarray}
\ln\left(\frac{k_{\rm{max}}}{k_*}\right)=\frac{1-n_s}{\alpha_s} .
\end{eqnarray}
{}For EI to be prevented, we must simply enforce that the maximum of
the curvature power spectrum is less than $P_{\rm{FDR}}$, 
\begin{eqnarray}
    \ln P(k_{\rm{max}})=\ln P_* -\frac{(1-n_s)^2}{\alpha_s} <\ln
    P_{\rm{FDR}}.
\end{eqnarray}
This is equivalent to an upper bound on the running $\alpha_s$ of
\begin{eqnarray}
    \alpha_s<
    \frac{(1-n_s)^2}{\ln\left(P_*/P_{\rm{FDR}}\right)}.
    \label{eq:alpha_WI_bound}
\end{eqnarray}
Notice that this upper bound is generically looser in WI compared to
the standard CI scenario, by a factor $\ln
P_*/\ln(P_*/P_{\rm{FDR}})$. As expected, for $Q=0$, i.e.
$P_{\rm{FDR}}=1$, this bound reduces to the CI result obtained in
Ref.~\cite{Kinney:2014jya}.  Given the updated CMB bounds on the
spectral index ($1-n_s<0.05$) and the amplitude of the scalar
perturbations ($P_*\simeq 2.1\times 10^{-9}$), the upper bound on the
running in the CI limit is
\begin{equation}
    \alpha_s< -1.3 \times 10^{-4} \,\,\,\,\,\,\,\,\,\, {\rm (CI)}. \label{eq:alpha_CI_bound}
\end{equation}
The bound on the running, generalized in the WI context, is therefore
\begin{equation}
    \alpha_s<- 1.3\times 10^{-4} \times \frac{\ln P_{*}}{\ln(P_{*}/P_{\rm{FDR}})}\,\,\,\,\,\,\,\,\,\, {\rm (WI)}, \label{eq:alpha_WI_bound}
\end{equation}
where $\ln
P_*/\ln(P_*/P_{\rm{FDR}})\leq1$ for $P_{\rm{FDR}}\geq1$. In short, for any non-zero value of the dissipation strength $Q$,\footnote{The values of $Q$
for the condition of Eq.~\eqref{eq:alpha_WI_bound} are chosen when the
FDR condition in Eq.~\eqref{eq:FDR} is an equality, i.e., at the
threshold when $\delta\phi_S=\delta\phi_D$.} the upper bound on the running moves to smaller negative values   than in the CI case, i.e. it gets easier to avoid EI. This difference is significant in the case of strong dissipation $(Q\gg 1$), and particularly for $b_G\geq 0$, which corresponds to a dissipation rate with a non-negative temperature dependence. For example,
if we set $Q = 100$ and $b_G = 6.52$, which corresponds to the value found for a dissipation rate
$\Upsilon\propto T^c$, with $c=3$~\cite{Montefalcone:2022jfw}, we obtain
\begin{equation}
    \alpha_s< -4.4 \times 10^{-5}, \qquad \text{$(Q=100, b_G=6.52)$}\label{eq:alpha_WI_MWI_bound}
\end{equation}
which is roughly a factor of $3$ smaller than what we found in the CI
case. 
Overall, the above discussion emphasizes one of the main takeaways of our work: in WI a tinier amount of negative running than was required in the CI framework 
is sufficient to prevent EI. This is particularly pertinent, as models featuring a relatively large negative running are quite generic in a WI scenario.

As an example, the Minimal Warm Inflation (MWI) model~\cite{Berghaus:2019whh} is a successful WI construction where the
inflaton is treated as an axionic field with a Chern-Simons coupling
to non-Abelian gauge fields. MWI is realized in the SDWI regime ($Q\gg1$) and for a dissipation rate cubic in temperature, i.e. $\Upsilon\propto T^3$; hence the upper bound on the running in Eq.~\eqref{eq:alpha_WI_MWI_bound} roughly applies. The running of the
spectral index associated with this model was recently computed in~\cite{Das:2022ubr}, with the result
\begin{equation}
\alpha_s^{\rm{MWI}}\simeq -2\times 10^{-4}, \label{eq:alpha_MWI_computed}
\end{equation}
which, according to the bound in Eq.~\eqref{eq:alpha_WI_MWI_bound}, is
sufficiently negative to prevent EI from ever occurring. Similar
values of  the running were also found in Ref.~\cite{Das:2022ubr} for
a variant of MWI which implements an inflaton potential of the runaway
type~\cite{Das:2020xmh} and leads also to $\alpha_s\simeq
-4\times10^{-3}$, which again is safely below the bound in
Eq.~\eqref{eq:alpha_WI_MWI_bound}.

A relatively large negative running
of the spectral index is also achieved for WI realizations in the WDWI
regime ($Q\ll1$), as previously shown in Ref.~\cite{Benetti:2016jhf},
where, for a wide range of inflaton potentials  and forms of the
dissipation rates, $\alpha_s\lesssim-10^{-4}$. A few examples of models that produce such large negative values for the running include a quartic and a hilltop potential, respectively for a dissipation rate cubic and linear in temperature.

It is important to
stress that the values of the running quoted here are also consistent
with the latest observational constraints from \textit{Planck},
Eq.~\eqref{eq:Planck_alphaS}, at the 1$\sigma$ level. This emphasizes another relevant takeaway of our work: within the context of WI, CMB data are consistent with a large
arrays of inflationary models that avoid the onset of EI, thereby
restoring a simpler picture of the Universe evolution.

Finally, it is important to emphasize that in the above analysis we assumed that the running of the spectral
index is constant, i.e. that the running of the running $\beta_s$ and
higher order terms ($\gamma_s$, $\delta_s$, etc.) are all set to
zero. As already pointed out in~\cite{Kinney:2014jya} this restriction
is not necessary: in order to prevent the onset of EI $\alpha_s$ does
not need to be constant, it simply needs to be sufficiently
negative. More precisely, if we take higher-order terms in the
spectral index to be non-zero, i.e. we take the scalar power spectrum to be of the form
\begin{equation}
    P(k)=P_*\left(\frac{k}{k_*}\right)^{n_s-1+\frac{\alpha_s}{2}\ln(k/k_*)+\frac{\beta_s}{6}\ln^2(k/k_*)+\frac{\gamma_s}{24}\ln^3(k/k_*)+\dots},
\end{equation}
the bound in
Eq.~\eqref{eq:alpha_WI_bound} to prevent eternal inflation
becomes:
\begin{equation}
\alpha_s+\frac{2}{3}\beta_s+\frac{1}{4}\gamma_s+\cdots <
\frac{(1-n_s)^2}{\ln\left(P_*/P_{\rm{FDR}}\right)},
\end{equation}
which is always satisfied as long as Eq.~\eqref{eq:alpha_WI_bound}
holds and
\begin{equation}
    |\alpha_s|>\frac{2}{3}|\beta_s|>\frac{1}{4}|\gamma_s|,\,\text{etc.}
\end{equation}
 This hierarchy, typical of CI constructions, is also generally
expected in the context of WI (see, e.g., Ref.~\cite{Das:2022ubr}), as
higher order-terms in the spectral index will generally be
proportional to larger powers of the slow-roll parameters $\epsilon_W$
and $\eta_W$.\footnote{Recall that $\epsilon_W$ and $\eta_W$ are,
respectively, equal to the CI slow-roll parameters $\epsilon_V$ and
$\eta_V$ divided by $(1+Q)$, to take into account the effect of
dissipation. According to this definition, the slow-roll approximation
in WI is valid as long as   $\epsilon_W,\eta_W \ll
1$~\cite{Berera:2008ar}. Thus higher orders of $\epsilon_W$, $\eta_W$
are generally suppressed.}

\section{Conclusions}
\label{conclusions}

In this paper, we have studied  the conditions that lead to the
so-called \textit{eternal inflation} (EI), with a particular focus to the
warm inflation (WI) framework, where thermal dissipation effects
significantly alter the inflaton dynamics relative to the standard
cold inflationary (CI) picture. Specifically, we have extended the work of
Ref.~\cite{Kinney:2014jya} and generalized the upper bound on the running of
the scalar spectral index $\alpha_s$ necessary to prevent EI, when also incorporating the effects of thermal dissipation.

We showed that, similarly to what was found in the context of CI, a
sufficiently negative running  on super-horizon scales is sufficient
to prevent EI from ever occurring.  Assuming a constant running, we
derived the upper bound for $\alpha_s$,
\begin{eqnarray}
    \alpha_s<- 1.3\times 10^{-4} \times \frac{\ln P_{*}}{
      \ln(P_{*}/P_{\rm{FDR}})}, \label{eq:Conclusion_1}
\end{eqnarray}
where $P_*$ is the amplitude of the scalar power spectrum (at the
relevant scale $k_*$ for CMB observations) and $P_{\rm{FDR}}$, defined in Eq.~\eqref{eq:P_FDR}, represents the upper limit on the scalar power spectrum beyond which a
fluctuations dominated regime (FDR) takes over, thereby causing the
onset of EI. In standard CI, $P_{\rm{FDR}}=1$, while in the context of
WI, $P_{\rm{FDR}}>1$ and specifically has a positive power-law
dependence on the strength of the dissipation $Q\equiv \Upsilon/(3H)$
(see for instance Eq.~\eqref{eq:EIC_SDwarm}). This means that in a WI setting, an even tinier amount of
negative running than the one that was required in the CI framework is sufficient to prevent EI from ever occurring. 

In addition, we provided some examples of physically motivated WI
constructions which are consistent with the CMB constraints on all of
the relevant primordial observables (such as the tensor-to-scalar
ratio $r$, the spectral index $n_s$, as well as the running
$\alpha_s$) and also satisfy the bound in
Eq.~\eqref{eq:Conclusion_1} i.e., they do not have EI. These include the Minimal Warm Inflation (MWI)~\cite{Berghaus:2019whh} model and its variants~\cite{Das:2020xmh} in the strongly dissipative regime ($Q\gg 1$)~\cite{Das:2022ubr}; as well as a quartic or a
hilltop potential, respectively for a dissipation rate cubic and linear in temperature, in the weakly dissipative regime ($Q\ll 1$)~\cite{Benetti:2016jhf}. This emphasizes one of the main takeaways
of this work, which is that WI provides a natural framework to prevent
EI from occurring, while still remaining consistent with the stringer
constraints from CMB observations.

{}Finally, as already pointed out in~\cite{Kinney:2014jya}, we note
that in a more realistic scenario where the running of the running
$\beta_s$, and higher order terms ($\gamma_s$, $\delta_s$, etc.) are
non-zero,  as long as the scalar power spectrum remains smaller than
$P_{\rm{FDR}}$, it is still the case that EI does not occur. This can
be achieved as long as Eq.~\eqref{eq:Conclusion_1} holds and
$|\alpha_s|>\frac{2}{3}|\beta_s|>\frac{1}{4}|\gamma_s|$, etc. This is
an hierarchy generally  expected in the context of WI, as higher
order-terms in the spectral index will be proportional to larger
powers of the slow-roll parameters $\epsilon_W$ and
$\eta_W$~\cite{Das:2022ubr}.

\begin{acknowledgments}
The authors thank Luca Visinelli for the useful comments on a preliminary version of this manuscript.  R.O.R. acknowledges financial support  by research grants from
Conselho Nacional de Desenvolvimento Cient\'{\i}fico e Tecnol\'ogico
(CNPq), Grant No. 307286/2021-5, and from Funda\c{c}\~ao Carlos Chagas
Filho de Amparo \`a Pesquisa do Estado do Rio de Janeiro (FAPERJ),
Grant No. E-26/201.150/2021.  K.F.\ is Jeff \& Gail Kodosky Endowed Chair in Physics at the University of Texas at Austin, and K.F.\ and G.M.\ are grateful for support via this Chair. K.F.\ and G.M.\ acknowledge support by the U.S.\ Department of Energy, Office of Science, Office of High Energy Physics program under Award Number DE-SC-0022021 as well as support from the Swedish Research Council (Contract No.~638-2013-8993).  

\end{acknowledgments}

\bibliographystyle{JHEP}
\bibliography{bibl.bib}

\end{document}